\documentclass[
12pt,
pra,
reprint,
a4paper,
longbibliography,
nofootinbib,
nobibnotes,
superscriptaddress,
]{revtex4-2}
\usepackage[utf8]{inputenc}
\usepackage[T1]{fontenc}

\usepackage{amsmath,amssymb,amsthm,amsfonts,mathtools}
\usepackage{dsfont}
\usepackage{braket}

\usepackage[usenames,dvipsnames,table]{xcolor}

\theoremstyle{definition}

\def\ket#1{|#1\rangle}

\def\ketbra#1#2{|#1\rangle\langle #2 |}

\def\id{\mathbb{I}}%

\newcommand{\new}[1]{{\color{black} #1}}

\usepackage[colorlinks=true,citecolor=blue,urlcolor  = purple]{hyperref}

\begin{document}

\title{
No-go theorem based on incomplete information of Wigner about his friend
}
\author{Zhen-Peng Xu}
\email{zhen-peng.xu@ahu.edu.cn}
\affiliation{School of Physics and Optoelectronics Engineering, Anhui University, 230601 Hefei, People’s Republic of China}
\affiliation{Naturwissenschaftlich--Technische Fakult\"{a}t,
Universit\"{a}t Siegen, 57068 Siegen, Germany}
\author{Jonathan Steinberg}
\email{steinberg@physik.uni-siegen.de}
\affiliation{Naturwissenschaftlich--Technische Fakult\"{a}t,
Universit\"{a}t Siegen, 57068 Siegen, Germany}
\author{H. Chau Nguyen}
\email{chau.nguyen@uni-siegen.de}
\affiliation{Naturwissenschaftlich--Technische Fakult\"{a}t,
Universit\"{a}t Siegen, 57068 Siegen, Germany}
\author{Otfried G\"uhne}
\email{otfried.guehne@uni-siegen.de}
\affiliation{Naturwissenschaftlich--Technische Fakult\"{a}t,
Universit\"{a}t Siegen, 57068 Siegen, Germany}

\begin{abstract}
The notion of measurements is central for many debates in quantum mechanics. One critical point is whether a measurement can be regarded as an absolute event, giving the same result for any observer in an irreversible manner. Using ideas from the gedankenexperiment of Wigner's friend it has been argued that, when combined with the assumptions of locality and no-superdeterminism, regarding a measurement as an absolute event is incompatible with the universal validity of quantum mechanics. We consider a weaker assumption: is the measurement event realised relatively to the observer when he only partially {\color{black}observed} the outcome. We proposed a protocol to show that this assumption putting in conjunction with the natural assumptions of no-superdeterminism and locality is also not compatible with the universal validity of quantum mechanics. 
\end{abstract}


\maketitle
\section{Introduction}
At the center of many orthodox interpretations
of quantum mechanics is the assumption that the action of measuring a 
quantity on a physical system creates its actual value 
\cite{jammer1974,mermin1985,marage1999}. This prompts to suggest 
that the action of creating a value of a measured quantity is an 
`absolute event' of this world, meaning that it is same for any 
observer and a process that cannot be reversed~\cite{brukner2017quantum,brukner2018no,bong2020strong,cavalcanti2018classical}. 
This perception of measurements in quantum mechanics has been 
highly debated. While it is supported by collapse models 
\cite{ghirardi1986,pearle1989} of the measurement process, 
it is not in line with other viewpoints which assume the 
universal validity of quantum mechanics \cite{zeh1970,zurek1982a,zurek2003a,arndt1999wave,monroe1996,davidovich1996,schlosshauer2005,anderson1993negotiating}. 
Indeed, assuming universality of quantum mechanics suggests to model the 
measurement process by a unitary dynamics, which in principle can always 
be reversed~\cite{schlosshauer2005}. As a result, the measurement may be 
undone and the value of the measured quantity can be erased, as if it never 
existed.

The subtleties of these different views are often discussed using a 
gedankenexperiment known as Wigner's friend \cite{wigner1995,deutsch1985}. In this scenario, the physicist Wigner has a friend in a box and this 
friend performs a measurement on a quantum system. For Wigner, the 
box undergoes a unitary evolution, but at a later point, Wigner may 
open the box and read out what his friend has measured. This, however, 
leads to a disagreement on when the absolute measurement event has 
taken place -- when the friend measured or when Wigner opened the box? 
Variations of this scenario have sparked interesting discussions recently \cite{frauchiger2018,brukner2017quantum,brukner2018no,brukner2020,bub2020,sudbery2019,bagio2021}, where sometimes even the consistency of quantum mechanics was questioned. 

\begin{figure}[t!]
  \centering
  \includegraphics[width=0.45\textwidth]{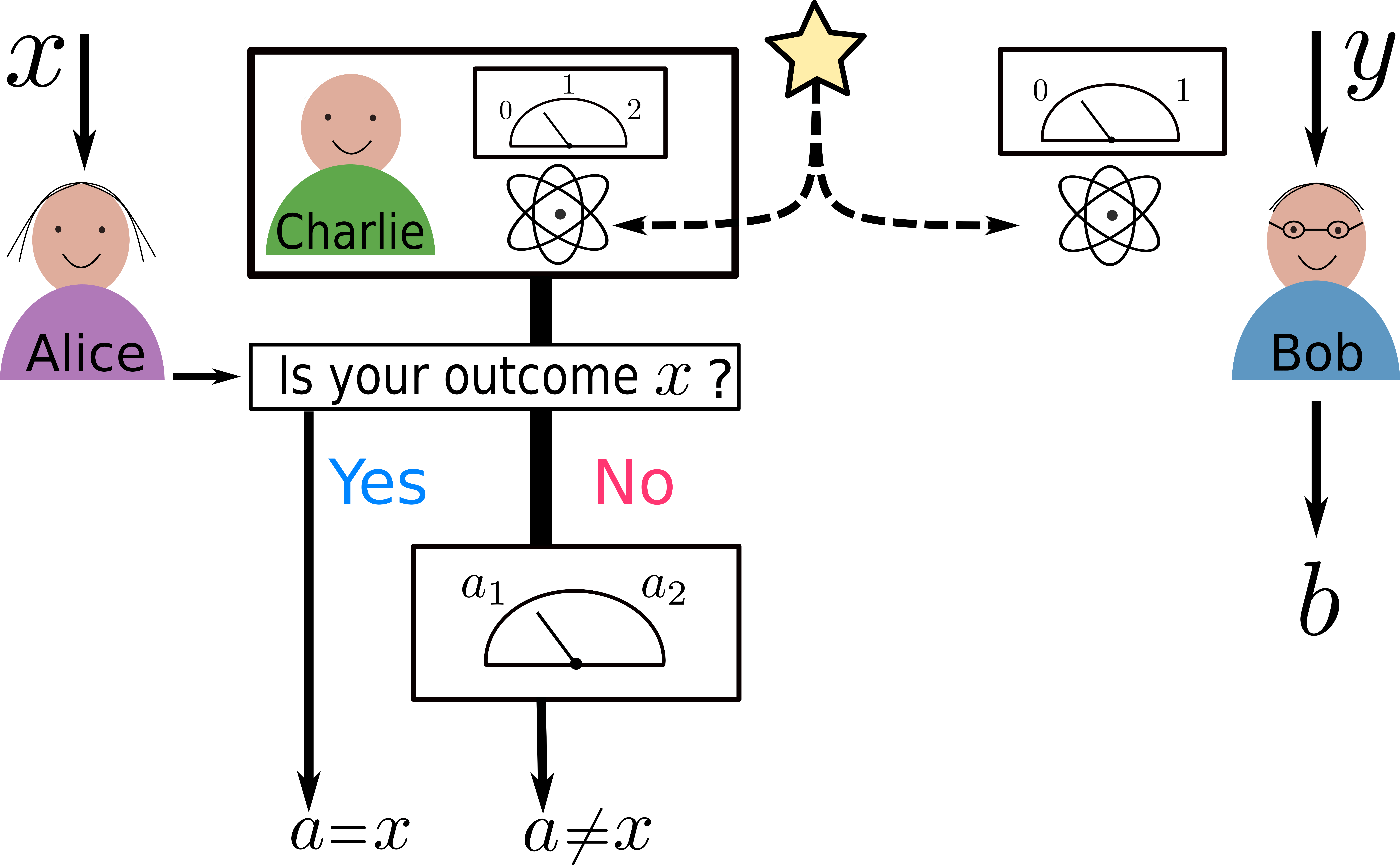}
  \caption{Illustration of the protocol. A pair of particles are distributed 
  to Charlie and Bob. Charlie, playing the role of Wigner's friend in a closed laboratory, carries out a measurement with outcomes $0$, $1$, $2$. Alice, playing 
  the role of Wigner, receives a signal $x \in \{0,1,2\}$ . She asks Charlie 
  if his outcome is $x$ or not. If it is $x$, Alice uses it as her output $a$. Otherwise, Alice continues to carry out a binary outcome measurement on the 
  system and the whole laboratory to obtain an outcome $a \in \{0,1,2\}$ but 
  $a \neq x$. Bob receives a signal $y\in \{0,1\}$ and performs
  correspondingly one of two possible binary outcome measurements on his particle 
  and output the outcome $b$.}
  \label{fig:friend}
\end{figure}

Still, the question whether the event of a measurement can be reversed 
or happens in an \emph{absolute sense}, is differently answered by different 
physicists. Consequently, this question should be answered with the help
of experiments, and not only theoretical considerations. Recently, 
Bong et al.~\cite{bong2020strong} provided a significant step in this
direction. They analyzed a so-called local friendliness (LF) scenario, 
where the assumption of absolute measurement events (AOE) is combined 
with other natural assumptions. In their protocol, a  measurement 
is carried out by Charlie, who {plays the role of Wigner's friend} in a closed laboratory controlled 
by Alice. Alice later can decide to either ask Charlie for the 
measurement outcomes, or simply reverse his measurement and perform 
an own measurement. The idea is that the \emph{persistence} of a 
measurement value which is implied by the absoluteness of the event, 
when combined with the assumptions of locality and no-superdeterminism{,}
puts strong constraints on the observable statistics, which can be 
violated if quantum mechanics is universally 
valid~\cite{brukner2018no,bong2020strong}. A proof-of-principle 
experiment has been also carried out showing the disfavour of the assumption of absoluteness of a measurement event~\cite{bong2020strong}.

\new{One may argue that assuming the absoluteness of a measurement with respect to every observer is a too strong assumption~\cite{workshop}. 
This is particularly clear under the viewpoint of universally valid quantum mechanics, in which the measurement can eventually be reversed, erasing any trace of existence of an outcome. This is the case notwithstanding Deutsch's proposal~\cite{deutsch1985} that Wigner can receive a message of the form `I see a definite outcome' from his friend without destroying the superposition of his friend. Indeed, if universal quantum mechanics is assumed, the writing action of Wigner's friend is also a unitary process. By the linearity of the unitary evolution, it is easy to show that such a unitary evolution would be uncorrelated with any state of the measurement device, be it in a definite state of a preferred basis or superpositions thereof.~\footnote{Indeed, the evolution is expected to map $\ket{i}\ket{\mbox{`blank'}}$ to $\ket{i}\ket{\mbox{`I see a definite outcome'}}$, where $\ket{i}$ is the state of the measurement device pointing to outcome $i$, and the second factor describes the state of the paper carrying the message from Wigner's friend. By linearity, this also implies that the unitary evolution maps $\ket{\psi}\ket{\mbox{`blank'}}$ to $\ket{\psi}\ket{\mbox{`I see a definite outcome'}}$ for \emph{any} state $\ket{\psi}$ of the measurement device.}}

\new{Therefore, we propose to investigate a weaker assumption than AOE: the persistence of the measurement event relative to the observer who has gained partial information about the measurement outcome. That is to say, an observer holding a coarse description of the outcome, is convinced of the existence of its fine description. We can refer to this assumption as `relative event by incomplete information (REII)'.}
We show that this assumption can also be rejected by the universal validity of quantum mechanics under the assumptions of no-superdeterminism and locality. The idea is to allow the measurement by \new{Wigner's friend} to have at least three outcomes, for example $a=0,1,2$; 
see Fig.~\ref{fig:friend}. 
In this scenario, it is possible 
for \new{Wigner} to ask \new{his friend} whether his measurement has yielded outcome in a {certain} coarse \emph{set}, say $\{0,2\}$, without accessing its fine, exact value.
\new{It is interesting to see that, unlike the original Wigner friend scenario discussed in Ref.~\cite{deutsch1985,brukner2017quantum,bong2020strong}, even when quantum mechanics is universally valid, in general, Wigner is not able to reverse the whole experiment of his friend; the coarse description of the outcome remains valid to him and his friend. {REII} assumes that this coarse record implies the existence of a fine-grain description of the outcomes, although it could have been forgotten by the friend. That quantum mechanics violates this assumption can be interpreted as: this coarse record is all to be there; a fine-grain description is still `unspeakable'.}
Our results show that not only the existence of measurement events is relative~\cite{brukner2020}, but the nature of the events themselves is defined by how the observer perceives the outcomes.

\section{The main protocol} \label{sec:main_protocol}
\new{To illustrate the idea, we consider here a minimal protocol.
Subtleties and further elaborations are to be discussed later in Sec.~\ref{sec:app4}.}
Consider Alice  and Bob sharing two particles at two different locations. 
Alice stores her particle in a laboratory, in which she has 
\new{Charlie playing the role of Wigner's friend}. In each run of the protocol, Charlie performs a measurement with three 
outcomes $c \in \{0,1,2\}$ on the particle. Then Alice receives a 
signal $x \in \{0,1,2\}$. Given the signal $x$, she asks 
Charlie if his measurement outcome is $c=x$. If this is the 
case, Alice outputs $a=x$, else she makes a binary outcome 
measurement on the particle and use the outcome to decide 
the output among $\{0,1,2\} $ with the constraint that 
$a \neq x$.
On the other hand, Bob receives a signal $y \in \{0,1\}$ in each run of the protocol, based 
on which he chooses one of two measurements to perform on his 
particle to get a binary outcome $b \in \{0,1\}$. Over many runs of the protocol, the collected 
data allows one to compute the statistics $p(a,b|x,y)$. 
In the following, we then derive explicit inequalities 
based on the assumption of the realisation of measurement event to the observer who partly gains the information about the outcomes (Alice in this case) combined with other natural assumptions. 
Still, the inequalities are 
violated in quantum mechanics, if its universal validity
holds.

\subsection{\new{Relative event by incomplete information (REII)}}


{Because  Alice has obtained part of the information about the outcomes of Charlie's measurement in each of the runs, REII directly implies that the value of $c$ is 
realized in every run of the protocol.} 
So, for every given $x$ and $y$ in each of the runs, {Alice can assume} a joint distribution $P(a,b,c|x,y)$ such that the observed data $p(a,b|x,y)$ is given by marginalizing 
the irrelevant outcome $c$,
\begin{equation}\label{eq:aoe}
p(a,b|x,y) = \sum_{c} P(a,b,c|x,y).
\end{equation}
Here and in the following we use the lower case letter $p$ to indicate 
that $p(a,b|x,y)$ is experimentally accessible, rather than being hypothetical like 
$P(a,b,c|x,y)$ with a capital $P$.~\footnote{\label{fn} Notice that, if we are to apply our assumption of { relative event by incomplete information} to the scenario in Ref.~\cite{bong2020strong}, the existence of $c$ is not implied when Alice does not inquire Charlie at all.} 
Since if $c=x$ then $a=x$, we have 
$P(a=x|c,x,y) = \delta_{xc}$. This means that the existence of $c$ can 
be consistently revealed when it is read. Noticing $P(a=x,b|c,x,y) = P(b|a=x,c,x,y) P(a=x|c,x,y)$, this then implies that
\begin{align}
P(a\!=\!x,b|c,x,y)  =
\delta_{xc} P(b|a\!=\!x,c,x,y)
=\delta_{xc} P(b|c,x,y).
\label{eq:aoe1}
\end{align}
It is unknown 
how to reject this thesis purely by its own. However, when combined with two other 
seemingly natural assumptions: {freedom of choice} (or {no-superdeterminism}) 
and {locality}, it has a stringent constraint on the observable statistics 
$p(a,b|x,y)$.

\subsection{\new{Freedom of choice and locality}}

The assumption of the {\it freedom of choice} demands that the random inputs $x$ and $y$, 
are uncorrelated with any relevant variable in the experiment~\cite{wiseman2014}. In 
this context, this means that $x$ and $y$ are independent from the variable $c$, $P(c|x,y)=P(c)$.  One therefore can write $P(a,b,c|x,y)= P(a,b|c,x,y) P(c)$,
and hence
\begin{equation}\label{eq:foc}
p(a,b|x,y)= \sum_{c} P(a,b|c,x,y) P(c).
\end{equation}

{The freedom of choice is justified if Charlie makes the  measurement before Alice and Bob make the choices $x$ and $y$. This relied on the honesty of Charlie and the functionality of his device~\cite{bong2020strong}. 
Interestingly, part of this problem can also be addressed within our protocol; see Sec.~\ref{sec:app4}.}

The {\it locality assumption} implies that  Bob's measurement result $b$ 
does not depend on Alice's input $x$ and there is an analogue independence
between $a$ and $y$. Then
\begin{equation}\label{eq:locality}
P(a|c,x,y)=P(a|c,x) \mbox{ and } P(b|c,x,y)=P(b|c,y). 
\end{equation}
It should be emphasized that this notion of signal locality is weaker than the so-called
local causality~\cite{wiseman2014} and the probabilities $P(a|b,c,x,y)$ and 
$P(b|a,c,x,y)$ in general cannot be further simplified.

\new{We have deliberately used the locality notion from Ref.~\cite{bong2020strong}, resembling that of the local friendliness.
Had we used the stronger assumption such as local causality, Bell-like correlations were to be obtained~\cite{brukner2017quantum,brukner2018no,bong2020strong}. 
We emphasize that, in either case the assumption of REII is used instead of that of AOE. }


\subsection{Combining the assumptions}
The combination of these assumptions may, in accordance with the terminology
developed in Ref.~\cite{bong2020strong}, be called {\it local friendliness under 
incomplete information} (LFIC). Using this combination, we find that the probability distribution $p(a,b|x,y)$ must take the form
\begin{align}
&p(a,b|x,y) = \sum_{c} P_{\mathrm{NS}} (a,b|c,x,y) P(c), \label{eq:LF1}\\
&p(a=x,b|x,y) = \sum_{c} \delta_{xc} P(b|c,y) P(c),
\label{eq:LF2}
\end{align}
where $P_{\mathrm{NS}} (a,b|c,x,y)$ is a probability distribution constrained only by the assumption of locality and freedom of choice. On the one hand, Eq.~\eqref{eq:LF1} can be understood 
as a direct consequence of Eqs.~(\ref{eq:foc}, \ref{eq:locality}). 
On the other hand, Eq.~\eqref{eq:LF2} requires {the consistency of the measurement outcomes when they are read} in 
Eqs.~(\ref{eq:aoe1}, \ref{eq:LF1}).

It is interesting to see that Eq.~\eqref{eq:LF1} and Eq.~\eqref{eq:LF2} are a 
combination of the no-signaling model~\cite{popescu1994quantum,cirel1980quantum} 
and the local hidden variable model~\cite{brunner2014bell}. 
To mimic the scenario described in Ref.~\cite{bong2020strong}, we can adjust 
the protocol to reproduce the {original} LF model as follows: to respond to Alice's query, her 
friend just outputs the outcome of his measurement, which is also Alice's final 
output in the next step of the protocol. Then Eqs.~(\ref{eq:LF1}, \ref{eq:LF2}) 
reduce to $p(a,b|x,y) = \sum_{c} \delta_{ac} P(b|c,y) P(c)$ 
{\it for all} $x$, which is the original LF model. In this case, the LF model is also a local 
hidden variable model since there is effectively only a single measurement on 
Alice's side. 
{However, the LFIC correlation polytope is strictly larger
than LF polytope as mentioned in Footnote~\ref{fn}. 
In fact, here one can choose a cross-section in correlation space, such that  
both the LHV and LF correlations are empty, while the LFIC correlations 
are not.}

\subsection{The correlation polytope}
The correlations that are constrained by Eqs.~(\ref{eq:LF1}, \ref{eq:LF2})
form a polytope, to which there are $60$ facets~\cite{schrijver1998theory}. 
Among those, many coincide with the facets of the no-signaling polytope.
The remaining $32$ facets can be grouped into $4$ inequivalent classes by considering the symmetry between measurements and outcomes. 
Among these facets,  two classes only involve at most one 
measurement of Bob. Therefore, this makes no difference between the no-signaling  model and the local hidden variable model, thus they are omitted here. 
The other two representative inequalities of relevant classes are
\begin{align}
Z_{1} & :=  p(A_0=0,B_0=0) + p(A_1=1,B_1=0)
\nonumber \\ 
& - p(A_2=1,B_0=0) + p(A_2=1,B_1=1) \geq 0,
\label{eq:33322a}
\\
 Z_{2} & :=p(A_0=1,B_0=0) + p(A_0=2,B_1=1)
 \nonumber
 \\
 &+ p(A_1=1,B_0=1) - p(A_1=1,B_1=1) \geq 0.
 \label{eq:33322b}
\end{align}
Note that the inequality~\eqref{eq:33322a} includes $3$ measurements on Alice's side, 
but  inequality~\eqref{eq:33322b} includes only $2$ measurements on Alice's side.
The inequality~\eqref{eq:33322a} resembles the well-known Clauser-Horne-Shimony-Holt inequality, if the event $A_2=2$ is never realized, that is, whenever the query is 
$A_2$, Charlie always reply with a negative answer. A similar discussion applies 
for inequality~\eqref{eq:33322b} if the event $A_0=0$ is never realized.

\begin{figure}[t]
  \centering
  \includegraphics[width=0.45\textwidth]{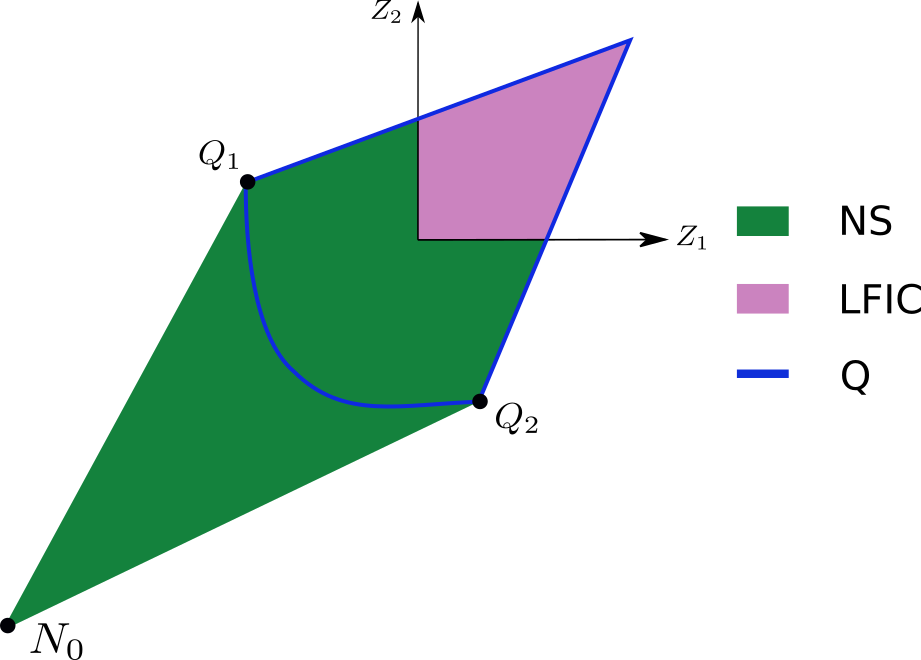}
  \caption{Cross sections of different models with a plane containing the points $Q_{1},Q_{2}$. The LFIC correlations (purple) are a subset of the NS correlations (green). The blue line bounds the set of correlations allowed by quantum theory within this section. {Note that all correlations on this plane cannot be described by a LHV or LF model.} 
  }
  \label{fig:cs322}
\end{figure}
In order to visualize the polytope, we compute a two-dimensional cross-section 
of the polytope with a plane defined by three points (see Fig.~\ref{fig:cs322}
and for details Appendix~A). 
To have a comparison, we have presented the cross-section with the polytope 
of the statistics which is constrained only by the locality condition, which 
is referred to as the non-signalling polytope. As expected, this polytope 
contains the polytope of the LFIC statistics.  Interestingly, 
the LF polytope, which coincides with the LHV polytope in this case, 
has no intersection with this plane.

\subsection{The quantum violation}
Finally, in  Fig.~\ref{fig:cs322}, we also present the set of correlations 
allowed by quantum mechanics if it is assumed to be universally valid. Its
boundary is computed using the Navascu\'es-Pironio-Ac\'{\i}n hierarchy up to 
the second level~\cite{navascues2008}. It clearly indicates that quantum mechanics 
violates the {LFIC model}. Accidentally, although this particular cross-section 
expresses a symmetry between $Z_1$ and $Z_2$ in Eq.~\eqref{eq:33322a} and Eq.~\eqref{eq:33322b}, the eventual inequivalence between them is revealed 
by the boundary of the quantum violation.

For a concrete physical realization, consider the case where Alice and Bob 
share a qutrit-qubit system in the state $\ket{\psi}=(\ket{00}+\ket{11})/\sqrt{2}$.
Alice's qutrit is stored in the lab and controlled by her friend Charlie. 
Charlie performs a measurement in the computational basis $\{\ket{0},\ket{1},\ket{2}\}$. 
Assuming the universal validity of quantum mechanics, the state of the 
joint system of Charlie and the qutrit-qubit system after the measurement 
is given by
\begin{equation}
\ket{\varphi}=\frac{1}{\sqrt{2}}    (\ket{0}_M \otimes \ket{00} + \ket{1}_M \otimes \ket{11}),
\label{eq:global_state}
\end{equation}
where the first term in the tensor products with subscript $M$ stands for Charlie's 
measurement device and the latter for the qutrit-qubit system. 
Receiving the random input $x$, Alice asks Charlie whether the outcome is $x$. 
In the case of a confirmation, Alice simply outputs $x$. In the case that it is not $x$, 
she performs measurements depending on the value of $x$ as follows. If $x=0$ or $x=1$, Alice can perform an arbitrary measurement and output the outcome. {This 
is because the events $A_0\neq 0$ and $A_1 \neq 1$ have not appeared in 
Eq.~\eqref{eq:33322a}, hence the choice of those two measurements does not 
affect the violation.} If $x=2$, she makes a unitary evolution to disentangle 
Charlie and his device from the qutrit-qubit system, bringing the latter back to $\ket{\psi}=(\ket{00}+\ket{11})/\sqrt{2}$ and effectively undoing Charlie's 
measurement. She then performs the measurement of $\sigma_x$ on the qutrit 
in the subspace spanned by $\ket{0}$ and $ \ket{1}$. Bob is making one of the two 
measurements $[-\sigma_x - \sigma_z]/\sqrt{2}$ and $[-\sigma_x + \sigma_z]/\sqrt{2}$. 
This setup in fact allows for a maximal violation of inequality~\eqref{eq:33322a} by 
$ Z_1=(1-\sqrt{2})/2 \approx 0.2071$, corresponding to the point $Q_1$ in 
Fig.~\ref{fig:cs322}. This violation is robust with respect to mixing
white noise to the state via $\varrho(p) = p \ketbra{\psi}{\psi} + (1-p)\openone/6$ 
as long as $ p \geq \frac{2}{17} \left(3 \sqrt{2}+1\right) \simeq 0.616781$.
A similar setup can be designed to attain the 
distribution $Q_2$ in Fig.~\ref{fig:cs322}, maximally violating  
inequality~\eqref{eq:33322b}. {More details are provided in Appendix~A.}

\section{Subtleties and an alternative protocol}\label{sec:app4}
In our main protocol, we have assumed that Charlie makes the measurement first and then Alice asks for the outcome.
In principle, this assumption does not hold if Charlie is not honest or the device does not function properly. For example, Charlie can wait for the inquiry from Alice and then implement the measurement, or Charlie can answer without referring to the exact measurement outcome. In principle, Alice can always open the box to check whether Charlie's answer is consistent with the outcome of the measurement device or not. Thus, we can simply assume that Charlie's answer is consistent with the outcome of the measurement device. This however cannot guarantee the no-superdeterminism assumption if the outcome of the measurement (which Charlie supposely implemented before) can be impacted by Alice's inquiry. Here we propose another protocol to fix this loophole.
That is, to make sure that Charlie's answer reflects the outcome of the measurement, which does not depend on Alice's input used in the statistics. 

For each run,
\begin{itemize}
  \item Charlie makes a measurement and obtains one of four outcomes $c\in \{0,1,2,3\}$;
  \item Alice receives a random number $t \in \{0,1, 2,3\} $ and inquiries Charlie whether $c$ equals $t$;
  \item Alice receives a random number $x \in \{0,1, 2,3\} $ and inquiries Charlie whether $c$ equals $x$;
\begin{itemize}
    \item If $c=x$, Alice outputs $a=x$;
    \item If $c\neq x$, Alice continues to make a measurement and obtains an outcome $a\in \{0,1,2,3\}\backslash\{x\}$;
\end{itemize}
  \item Bob receives the input $y\in \{0,1\}$, make a measurement and obtains an outcome in $b \in \{0,1\} $;
\end{itemize}
The statistics after many runs is collected to estimate $p(a,b|x,y)$.

According to this protocol, although $c$ could depend on $t$, it is independent of $x$ under the assumptions of {REII} and no-superdeterminism. 
In the special case that $c\neq t$ and $x\neq t$ for any fixed $t$, the LFIC model of the current protocol just reduces to our main protocol. As we have discussed already, this model cannot describe the statistics predicted by quantum mechanics.

\section{Discussion and conclusion} 
The violation of the inequalities~\eqref{eq:33322a} and~\eqref{eq:33322b} 
by quantum theory is due to the fact that the qutrit at Alice's side still 
maintains entanglement with Bob's qubit after that Charlie responses negatively 
to the question ``Is $x=2$''. This points towards a relevant discussion about 
the nature of the measurement process. Originally, according to von 
Neumann \cite{vonneumann1932}, a measurement of a degenerate observable leads 
to full decoherence in the entire eigenbasis of the observable, such that the 
post-measurement state is diagonal in this basis. It was then recognised by
Lüders \cite{lueders1950} that this is not the appropriate formulation. If a 
degenerate measurement is made, then according to the Lüders' rule the 
coherence in the degenerate subspaces is unaffected by that. In fact, the 
validity of Lüder's rule has recently been observed experimentally \cite{pokorny2020}.

Our analysis of the assumption of 
{relative event by incomplete information (REII)} 
also highlights the difference between the viewpoints
of von Neumann and Lüders.  If we treat Alice's query about $i$ and Charlie's 
measurement as a single measurement $\tilde{c}_i$ implemented and read by 
Alice, it constitutes a dichotomic degenerate measurement on the qutrit. 
The post-measurement state following the Lüders' rule still has quantum 
coherence in a $2$-dimensional subspace, and thus entanglement with a 
remote party can still remain. In comparison, we see that von Neumann's 
rule is similar to the assumption of \new{REII}.

The violation of our inequalities can also be seen as related to an 
interesting remark by Peres in connection to the concept of
contextuality \cite{Peres2002,peres1978,budroni2022kochen}. 
There, in order to justify the contextuality assumption, he argued 
that it is essential that for a three-outcome measurement, Alice 
can construct a device to measure whether the system gives some 
outcome, e.g., $2$, and then, at a later stage or even in a different 
lab, decide how to complete the measurement.

While the main protocol presented in Sec.~\ref{sec:main_protocol} was of a particularly simple form, we stress that our findings
also hold for related scenarios that are more sophisticated; see the analysis in Sec.~\ref{sec:app4}. Another interesting extension would be to consider this quantum correlation sets with incomplete information 
beyond bipartite cases.
The investigation of a multipartite all-versus-nothing-like proof such as the one discussed in 
Ref.~\cite{brukner2018no} under incomplete information would also be very interesting.

Finally, asking for an experimental test would require the manipulation of
an entire lab. This brings us back to Bell's famous question \cite{bell2004}: What exactly qualifies some physical systems to play the role of `measurer'? Does Charlie's lab need to contain a 
physicist with a PhD? Still, proof-of-principle demonstrations, in the 
spirit of Ref.~\cite{bong2020strong} would be highly desirable.


{\it Acknowledgments.---}
The authors would like to thank Eric G. Cavalcanti and Howard Wiseman for critical comments on the early version of this work. 
Discussions with Fabian Bernards, Matthias Kleinmann were fruitful.  
We acknowledge Ad\'an Cabello for support 
of the discussions, and the University of Siegen for enabling our computations through the OMNI cluster. 
This work was supported by the Deutsche Forschungsgemeinschaft (DFG, German
Research Foundation, project numbers 447948357 and 440958198), the Sino-German Center for Research Promotion (Project M-0294), the ERC (Consolidator Grant 683107/TempoQ) and the German Ministry of Education and Research (Project QuKuK, BMBF Grant No. 16KIS1618K).
Z.P.X. acknowledges support from the Humboldt foundation. 
J.S. acknowledges support from the House of Young Talents of the University of Siegen.
{H.C.N. acknowledges the Unitary fund for supporting his attendance at the Wigner's friend workshop, and the participants for exciting discussions.}
\onecolumngrid

\appendix
\setcounter{equation}{9}
\section{Details of the main protocols}
\subsection{Details of the polytopes of correlations} \label{Ap:01}
The set of correlations $p(a,b|x,y)$ defined by Eqs.~(5) and (6) in the main text forms a polytope. 
The polytope can be fully characterized by its facets, which are described by inequalities.
The inequalities are derived from the definition Eqs.~(5) and (6) using a linear programing solver~\cite{cdd}.
There are {60} such inequalities to the polytope, among which {32} do not coincide with the ones from the no-signalling (NS) polytope. 
These {32} inequalities can classified into $4$ inequivalent classes, which are not related by the permutation of measurements and outcomes.
The representatives of these $4$ inequivalent classes are
\begin{align}
& p(A_0=0,B_0=0) + p(A_1=1,B_1=0) - p(A_2=1,B_0=0) + p(A_2=1,B_1=1) \geq 0,\label{eq:class1}\\
& p(A_0=1,B_0=0) + p(A_0=2,B_1=1) + p(A_1=1,B_0=1) - p(A_1=1,B_1=1) \geq 0,\label{eq:class2}\\
& p(A_0=0,B_0=1) + p(A_1=1,B_0=1) - p(A_2=1,B_0=1) \geq 0,\label{eq:class3}\\
&p(A_0=1,B_0=1)+p(A_0=2,B_0=1)-p(A_1=1,B_0=1) \geq 0,\label{eq:class4}
\end{align}
Additionally, the polytope is also constrained by three {inequivalent} hyperplanes {which are not from the NS polytope}
\begin{align}
&-p(A_0\neq 0)+p(A_1=1)+p(A_2=2) = 0,\label{eq:class5}\\
&-p(A_0\neq 0,B_0=1)+p(A_1=1,B_0=1)+p(A_2=2,B_0=1) = 0,\label{eq:class6}\\
&-p(A_0\neq 0,B_1=1)+p(A_1=1,B_1=1)+p(A_2=2,B_1=1) = 0.\label{eq:class7}
\end{align}
It turns out that, the inequalities in {from Eq.~\eqref{eq:class3} to Eq.~\eqref{eq:class7}} also hold if quantum mechanics is assumed.
Hence, only the inequalities in Eq.~\eqref{eq:class1} and Eq.~\eqref{eq:class2} are nontrivial.

\subsection{Details of the cross-section and quantum violation}\label{sec:cs}
We use three points $N_0$, $Q_1$, $Q_2$ to determine the plane for the cross-section in Fig. 2. 
The point $N_0$ is obtained in the NS model, the points $Q_1, Q_2$ can be achieved in quantum theory. In fact, $Q_1$ is one optimal solution for inequality~(7) in the main text, $Q_2$ is an optimal solution for inequality~(8) in the main text in quantum theory. The exact values of those three points are collected in Table~\ref{tab:points}. 

\begin{table}[!h]
\begin{tabular}{c|c|cc|cc|cc}
\hline\hline
$N_0$& $\id$ & $A_0=1$ & $A_0=2$ & $A_1=1$ & $A_1=2$ & $A_2=1$ & $A_2=2$\\ \hline
$\id$ & 1 & 1/2 & 1/2 & 1/2 & 0 & 1/2 & 1/2\\  \hline
$B_0 = 1$ & 1/2 & 0 & 1/2 & 0 & 0 & 0 & 1/2\\ \hline
$B_1 = 1$ & 1/2 & 0 & 1/2 & 1/2 & 0 & 0 & 1/2\\ \hline\hline
$Q_1$ & $\id$ & $A_0=1$ & $A_0=2$ & $A_1=1$ & $A_1=2$ & $A_2=1$ & $A_2=2$\\ \hline
$\id$ & 1 & 1/2 & 0 & 1/2 & 0 & 1/2 & 0\\  \hline
$B_0 = 1$ & 1/2 & $\alpha$ & 0 & $\alpha$ & 0 & $\alpha$ & 0\\ \hline
$B_1 = 1$ & 1/2 & $\beta$ & 0 & $\beta$ & 0 & $\alpha$ & 0\\
\hline\hline
$Q_2$ & $\id$ & $A_0=1$ & $A_0=2$ & $A_1=1$ & $A_1=2$ & $A_2=1$ & $A_2=2$\\ \hline
$\id$ & 1 & 1/2 & 1/2 & 1/2 & 0 & 0 & 1/2\\  \hline
$B_0 = 1$ & 1/2 & $\beta$ & $\alpha$ & $\alpha$ & 0 & 0 & $\alpha$\\ \hline
$B_1 = 1$ & 1/2 & $\beta$ & $\alpha$ & $\beta$ & 0 & 0 & $\alpha$\\
\hline\hline
\end{tabular}
\caption{The probability distributions $N_0$, $Q_1$, $Q_2$, where $\alpha = (\sqrt{2}-1)/4\sqrt{2}$, $\beta = (\sqrt{2}+1)/4\sqrt{2}$.}\label{tab:points}
\end{table}

The point $Q_1$ can be obtained using the state $\frac{\ket{00}+\ket{11}}{\sqrt{2}}$
and the measurement directions
\begin{align}
    &\ket{A_0=2} = \ket{A_1=2} = \ket{A_2=2} = \ket{2},\\
    &\ket{A_0=0} = \ket{A_1 = 0} = \ket{0},\\
    &\ket{A_0=1} = \ket{A_1 = 1} = \ket{1},\\
    &\ket{A_2=0} = \ket{+}, \ket{A_2 = 1} = \ket{-},\\
    &B_0 = \frac{-\sigma_x-\sigma_z}{\sqrt{2}}, B_1 = \frac{-\sigma_x+\sigma_z}{\sqrt{2}}.
\end{align}

The point $Q_2$ can be obtained using the state $\frac{\ket{00}+\ket{11}}{\sqrt{2}}$ and
the measurement directions
\begin{align}
    &\ket{A_0=0} = \ket{A_1=2} = \ket{A_2=1} = \ket{2},\\
    &\ket{A_0=1} = \ket{A_2 = 0} = \ket{0},\\
    &\ket{A_0=2} = \ket{A_2 = 2} = \ket{1},\\
    &\ket{A_1=0} = \ket{-}, \ket{A_1 = 1} = \ket{+},\\
    &B_0 = \frac{\sigma_x-\sigma_z}{\sqrt{2}}, B_1 = \frac{-\sigma_x-\sigma_z}{\sqrt{2}}.
\end{align}

\twocolumngrid
\bibliography{bib}

\end{document}